\documentclass[aps,prl,twocolumn,superscriptaddress]{revtex4}
\usepackage{epsf,graphicx}
\usepackage{amssymb}
\usepackage{amsmath}
\usepackage{latexsym,bm,array,amsfonts,multirow}
\usepackage{color}
\usepackage{ulem}
\usepackage{booktabs}
\usepackage{array}

\begin{document}

\title{Fermi liquid and isotropic superconductivity of Hund scenario for bilayer nickelates}
\author{Jiangfan Wang}
\email[]{jfwang@hznu.edu.cn}
\affiliation{School of Physics, Hangzhou Normal University, Hangzhou, Zhejiang 311121, China}
\author{Yi-feng Yang}
\email[]{yifeng@iphy.ac.cn}
\affiliation{Beijing National Laboratory for Condensed Matter Physics and Institute of
Physics, Chinese Academy of Sciences, Beijing 100190, China}
\affiliation{School of Physical Sciences, University of Chinese Academy of Sciences, Beijing 100049, China}
\affiliation{Songshan Lake Materials Laboratory, Dongguan, Guangdong 523808, China}
\date{\today}

\begin{abstract}
Recent experiments on bulk and thin film bilayer nickelate high-$T_c$ superconductors urge for clarification of their pairing mechanism. Debates exist on whether the hybridization or the Hund's coupling between the nickel $d_{x^2-y^2}$ and $d_{z^2}$ orbitals plays a primary role in driving the superconductivity. Here, we study the Hund scenario and make comparisons with the hybridization scenario using the same dynamic Schwinger boson approach. Our calculations reveal several key features of the Hund-driven superconductivity, including an isotropic $s$-wave gap, a lower maximum $T_c$, and Fermi liquid normal states, that differ from the hybridization-driven mechanism. We attribute these differences to their distinct low-energy dynamics. Comparison with recent experiments suggests that the Hund scenario alone is not enough to explain the bilayer nickelate superconductivity in both bulk and thin films.         
\end{abstract}
\maketitle

The discovery of bilayer Ruddlesden-Popper nickelate superconductor La$_{3}$Ni$_2$O$_7$ with a transition temperature $T_c \approx 80$ K under pressure has garnered significant attention \cite{MWang2023Nature,JGCheng2023,HQYuan2024,ZChen2024,JGCheng2024Nature,HHWen2024,XJZhou2024,LShu2024,DLFeng2024,MWang2024,MWang2024b,LYang2024,Shen2025}. More recently, superconductivity with $T_c \approx 40$ K was discovered in the compressively strained (La,Pr)$_{3}$Ni$_2$O$_7$ \cite{Hwang2025,Chen2025,Hwang2025b,QKXue2025,ZXShen2025,Hwang2025FL,He2025pseudogap,HHWen2025,ZYChen2025} and (La,Sr)$_{3}$Ni$_2$O$_7$ \cite{Nie2025} thin films at ambient pressure. Both host dominant nickel $3d_{z^2}$ and $3d_{x^2-y^2}$ bands near the Fermi level \cite{DXYao2023,Dagotto2023,Werner2023,Leonov2023,Eremin2023,YYCao2024,Eremin2024,MWang2023Nature,QKXue2025,ZXShen2025,DXYao2025film,WQChen2025,HHWen2024,XJZhou2024}, but exhibit significantly different $T_c$ and normal state properties: the bulk La$_3$Ni$_2$O$_7$ displays linear-in-$T$ resistivity above the superconducting dome under high pressure \cite{HQYuan2024,Shen2025}, while the thin films show Fermi liquid (FL) transport at ambient pressure \cite{Hwang2025FL}. While density functional theory calculations predict $d_{z^2}$ hole pockets for the high pressure phase of La$_3$Ni$_2$O$_7$ \cite{MWang2023Nature,DXYao2023}, angle-resolved photoemission spectroscopy (ARPES) measurements of (La,Pr)$_{3}$Ni$_2$O$_7$ thin films show contradictory $d_{z^2}$ band position relative to the Fermi level \cite{QKXue2025,ZXShen2025}. These experimental facts provide important clues to clarify the true pairing mechanism among all different  proposals.

Theoretically, most studies are based on certain types of bilayer two-orbital Hubbard model or $t$-$J$ model, and employ either weakly or strongly correlated approaches  \cite{WQChen2025,JPHu2025,QHWang2025,JPHu2025PRL,FYang2023,QHWang2023,YFYang2023,Wang2025,QQin2023,GMZhang2023,WWu2025,CJWu2024PRL,CJWu2024,CJWu2025,GSu2024,GSu2025,YHZhang2023,Bohrdt2024,DXYao2023tJ,YZYou2023,ZYWeng2024,ZYLu2024,Kuroki2024,WLi2024,KJiang2024,WKu2024,TXiang2024}. The weak-correlation theories treat the Fermi surface nesting and its associated spin density wave fluctuations as the origin of the pairing glue, but remain debated on whether it is sufficient to explain the high $T_c$ without resorting to electron-phonon interaction \cite{QHWang2025,JPHu2025PRL}. By contrast, the strong-correlation theories regard the interlayer superexchange interaction of the (nearly) half-filled $d_{z^2}$ electrons as the primary interaction for the pairing, but are divided as to whether the hybridization \cite{YFYang2023,Wang2025,QQin2023,GMZhang2023,WWu2025} or the Hund's coupling \cite{CJWu2024PRL,CJWu2024,CJWu2025,GSu2024,GSu2025,YHZhang2023,Bohrdt2024} play a key role in transmitting the pairing interaction to itinerant $d_{x^2-y^2}$ electrons for the superconducting phase coherence. 

In the hybridization scenario described by a bilayer two-orbital $t$-$V$-$J$ model \cite{YFYang2023,Wang2025,QQin2023}, where $t$, $V$, $J$ represents the intralayer $d_{x^2-y^2}$ hopping, hybridization, and $d_{z^2}$ interlayer superexchange, respectively, previous Schwinger boson calculations in the large $U$ limit have predicted a highly asymmetric superconducting dome upon $d_{z^2}$ hole doping, and a non-Fermi liquid (NFL) normal state around optimal $J$ (for maximum $T_c$) \cite{Wang2025}, in agreement with experiments of pressurized La$_{3}$Ni$_2$O$_7$ \cite{MWang2024b,HQYuan2024,Shen2025}. The maximum $T_c^\text{max}$ is found to be about $0.17J$ from mean-field type Schwinger boson calculations \cite{Wang2025}, and $0.05J$ from auxiliary-field Monte Carlo simulations where superconducting phase fluctuations are included \cite{QQin2023,QQin2025}. It has anisotropic $s^{\pm}$-wave gaps with nodes or minima along the zone diagonal on the $d_{x^2-y^2}$ Fermi surfaces ($\alpha$ and $\beta$) and an isotropic $s$-wave gap on the $d_{z^2}$ Fermi surface ($\gamma$), consistent with recent experiments of La$_2$PrNi$_2$O$_7$ \cite{HHWen2025} and (La,Pr,Sm)$_3$Ni$_2$O$_7$ thin films \cite{ZYChen2025}. On the other hand, the Hund scenario is often studied based on some type of effective single-band $t$-$J$ model of $d_{x^2-y^2}$ electrons in the infinite Hund's coupling limit \cite{CJWu2024PRL,GSu2024,YHZhang2023,Bohrdt2024}. Despite of some recent work focusing on finite Hund's coupling \cite{CJWu2025,GSu2025}, a comprehensive study of its phase diagram, $T_c^\text{max}$, superconducting and normal state properties based on a bilayer two-orbital model is highly demanded for a faithful comparison with recent experiments and other theories.

In this work, we study a bilayer two-orbital $t$-$J_H$-$J$ model using the dynamic Schwinger boson approach and focusing on the interplay between the inter-orbital Hund's coupling ($J_H$) and the $d_{z^2}$ interlayer superexchange ($J$). We found several important differences of the Hund scenario as summarized in Table \ref{tab1}: 1) Its pairing vertex of $d_{x^2-y^2}$ electrons exhibits no momentum dependence, suggesting a fully isotropic $s$-wave gap; 2) Its $T_c^\text{max}/J$ is about 40\% lower than the hybridization scenario; 3) Its $T_c$ is monotonically suppressed by the $d_{z^2}$ hole doping; 4) Its normal state is always a Fermi liquid irrespective of the Hund's coupling strength. Comparison of these results with recent experiments suggests that the Hund scenario alone is not enough to explain the superconductivity in both bulk and thin film bilayer nickelates.

\begin{table}[t]
	\centering
	\belowrulesep=0.65ex
	\aboverulesep=0.4ex
	\caption{Comparison between the Schwinger boson results of the $t$-$V$-$J$ model (hybridization scenario) in the large $U$ limit studied in Ref. \cite{Wang2025} and the $t$-$J_H$-$J$ model (Hund scenario) studied in this work. A factor of 0.3 is introduced for the $T_c^\text{max}$ to account for the phase fluctuations.}
	\label{tab1}
	\begin{tabular}{p{2.4cm}<{\centering} p{2.8cm}<{\centering} p{2.8cm}<{\centering}}\toprule[0.1em]
		   & $t$-$V$-$J$  & $t$-$J_H$-$J$   \\\midrule[0.05em] 
		Gap structure &  Anisotropic $s^\pm$ & Isotropic $s^\pm$ \\ 
		\specialrule{0em}{3pt}{3pt}
		$0.3T_c^\text{max}/J$ & $\approx 0.05$ & $\approx 0.03$   \\ 
		\specialrule{0em}{3pt}{3pt}
		$d_{z^2}$ metallization & Promotes SC & Suppresses SC \\
		\specialrule{0em}{3pt}{3pt}
		Normal state & NFL around optimal $J$ & FL for all $J$ \\\bottomrule[0.1em]
	\end{tabular}
\end{table}

\begin{figure}
	\begin{center}
		\includegraphics[width=8.6cm]{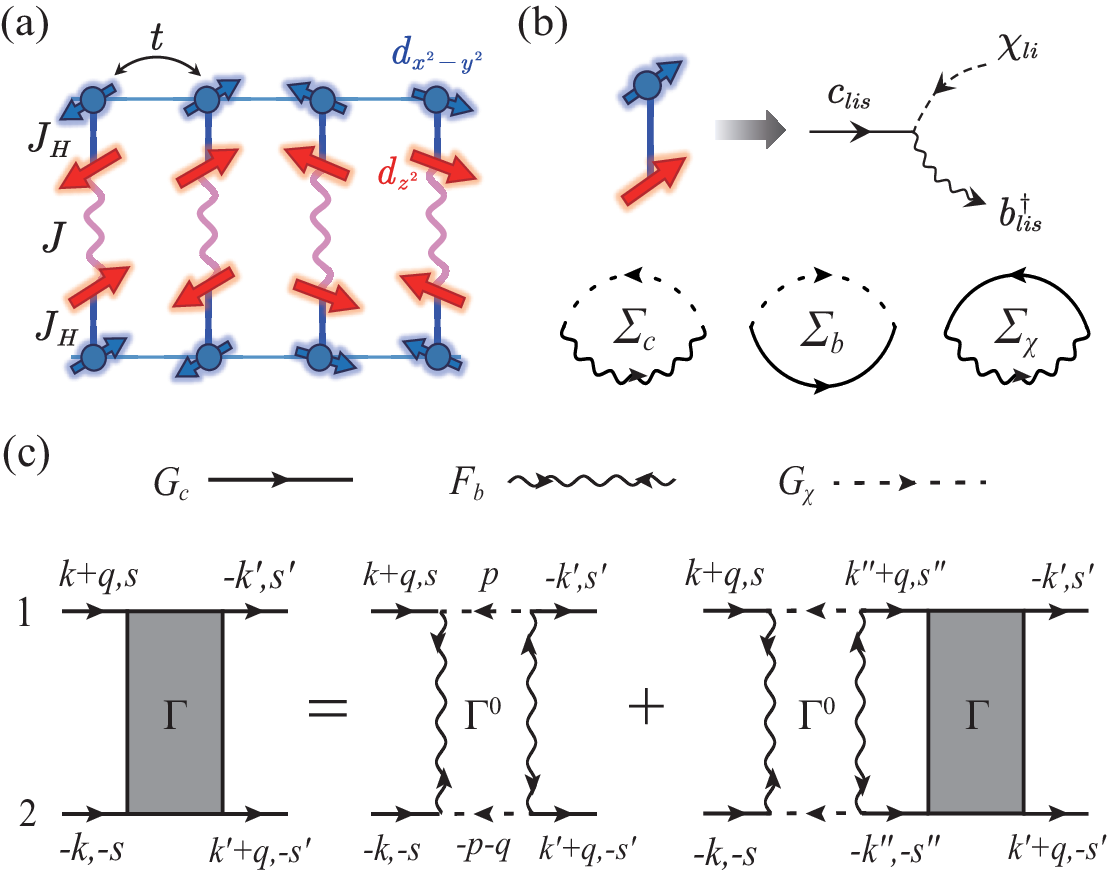}
	\end{center}
	\caption{(a) A schematic visualization of the bilayer two-orbital $t$-$J_H$-$J$ model along one axis of the two-dimensional square lattice. (b) An illustration of the Schwinger boson decomposition of the Hund's coupling term, and the Feynman diagrams for the self-energies of electrons, spinons, and holons. (c) A diagrammatic representation of the Bethe-Salpeter equation for the pairing vertex of $d_{x^2-y^2}$ electrons, where 1 and 2 denote the two layers and $k\equiv (\mathbf{k},i\omega_n)$, $k'\equiv (\mathbf{k}',i\omega_{n'})$, $q\equiv (\mathbf{q},i\nu_l)$, with $\omega_n$ ($\nu_n$) being the fermionic (bosonic) Matsubara frequency.}
	\label{fig:Fig1}
\end{figure}

\textit{Model---}We consider the effective $t$-$J_H$-$J$ model defined on a bilayer square lattice, 
\begin{eqnarray}
H&=&-\sum_{lijs}(t_{ij}+\mu\delta_{ij})c_{lis}^{\dagger}c_{ljs}-J_H\sum_{li}\bm{S}_{li}\cdot \bm{s}_{li} \label{eq:H} \nonumber\\
&&+J\sum_{i}\bm{S}_{1i}\cdot\bm{S}_{2i}, 
\end{eqnarray}
where $c_{lis}^\dagger$ creates a $d_{x^2-y^2}$ electron on site $i$ of the $l$-th ($l=1,2$) layer with spin $s=\pm 1$, $\bm{s}_{li}=\frac{1}{2}\sum_{ss'}c_{lis}^\dagger \boldsymbol{\sigma}_{ss'}c_{lis'}$ is its spin density operator, $\bm{S}_{li}$ denotes the local $d_{z^2}$ spin, $t_{ij}=t$ is the in-plane nearest neighbor hopping amplitude, and $\mu$ is the chemical potential of $d_{x^2-y^2}$ electrons. The above Hamiltonian is schematically represented in Fig. \ref{fig:Fig1}(a). To focus on the effect of the Hund's coupling and its interplay with the interlayer superexchange $J$, we have ignored other factors such as the relatively weaker intralayer superexchange \cite{DLFeng2024,MWang2024} and the Coulomb repulsion in the quarter-filled $d_{x^2-y^2}$ orbital.

\textit{Schwinger boson approach---}We rewrite the $d_{z^2}$ spin operator using the Schwinger boson representation, $\bm{S}_{li}=\frac{1}{2}\sum_{ss'}b_{lis}^\dagger \boldsymbol{\sigma}_{ss'} b_{lis}$, where the  spinon number constraint $n_{li}^b\equiv\sum_{s}b_{lis}^\dagger b_{lis}=1$ is implemented by a Lagrange multiplier term  $\sum_{li}\lambda_{li}\left(n_{li}^b-1\right)$. We then decompose the superexchange term as
\begin{eqnarray}
	J\boldsymbol{S}_{1i}\cdot \boldsymbol{S}_{2i}\rightarrow\Delta_i\sum_s s b_{1is}^\dagger b_{2i,-s}^\dagger+c.c.+\frac{2|\Delta_i|^2}{J}, \label{eq:HJ}
\end{eqnarray}
where  $\Delta_i$ is the interlayer valence bond field. We further take the mean-field approximation, $\Delta_{i}=\Delta$ and $\lambda_{li}=\lambda$, which are determined by the self-consistent equations $\Delta=-\frac{TJ}{2\mathcal{N}_s}\sum_{\mathbf{k}s}s\langle b_{1\mathbf{k}s}b_{2,-\mathbf{k},-s}\rangle$ and  $1=\frac{1}{2\mathcal{N}_s}\sum_{li}\langle n_{li}^b\rangle\equiv n_b$. The Hund's coupling term is factorized by introducing a fermionic holon field $\chi_{li}$,
\begin{eqnarray}
	-J_H\bm{S}_{li}\cdot \bm{s}_{li}\rightarrow & & \sum_s b_{lis}^\dagger c_{lis}\chi_{li}+c.c.-\frac{2|\chi_{li}|^2}{J_H} \notag \\
	& &+\frac{J_H}{4}\sum_s c_{lis}^\dagger c_{lis}, \label{eq:Hv}
\end{eqnarray}  
which leads to a three-particle interacting vertex shown in Fig. \ref{fig:Fig1}(b). The above decomposition applies to both ferromagnetic (Hund) and antiferromagnetic (Kondo) couplings \cite{Wang2020,Wang2021,Wang2022a,Wang2022b}. The resulting action is similar to that of the $t$-$V$-$J$ model \cite{Wang2025}, with a major difference in that the $\chi_{li}$ field introduced here possesses no intrinsic short-time dynamics. However, it can acquire long-time dynamics through the interacting vertex and self-energies shown in Fig. \ref{fig:Fig1}(b). The Green's functions depend self-consistently on the self-energies and are solved together with the mean-field variables $\Delta$ and $\lambda$ \cite{supp}.

\textit{Cooper instability---}We study the interlayer pairing instability of $d_{x^2-y^2}$ electrons by calculating the pairing vertex $\Gamma_{ss'}(k,k',q)$ using the Bethe-Salpeter equation shown in Fig. \ref{fig:Fig1}(c) \cite{Wang2025,supp}. The ``bare'' pairing vertex is momentum independent and given by
\begin{align}
	\Gamma_{ss'}^0&(i\omega_n,i\omega_{n'},i\nu_l)=\frac{ss'}{\beta}\sum_{m}F_b(i\omega_m+i\omega_n+i\nu_l)^* \label{eq:Gamma0} \notag \\
	&\times F_b(i\omega_m-i\omega_{n'})G_\chi(i\omega_m)G_\chi(-i\omega_m-i\nu_l),
\end{align}
where
$F_b(k)=-s\langle b_{1ks}b_{2,-k,-s}\rangle$  is the spinon anomalous Green's function. Consequently, for the static uniform solution ($q=0$), the total pairing vertex can be simplified to $\Gamma_{ss'}(k,k')=ss'\tilde{\Gamma}(i\omega_n,i\omega_{n'})$, which is also independent of momentum \cite{supp}. Thus, the superconducting gap is fully isotropic, in contrast to the hybridization induced anisotropic $s$-wave gap with nodes or minima along the zone diagonal on $\alpha$ and $\beta$ Fermi surfaces \cite{Wang2025,QQin2023}. $T_c$ is determined by the divergence of the real part of $\tilde{\Gamma}(i\omega_n,i\omega_{n'})$ at the minimal Matsubara frequency $|\omega_n|=|\omega_{n'}|=\pi T$.

\begin{figure}[t]
	\begin{center}
		\includegraphics[width=8.7cm]{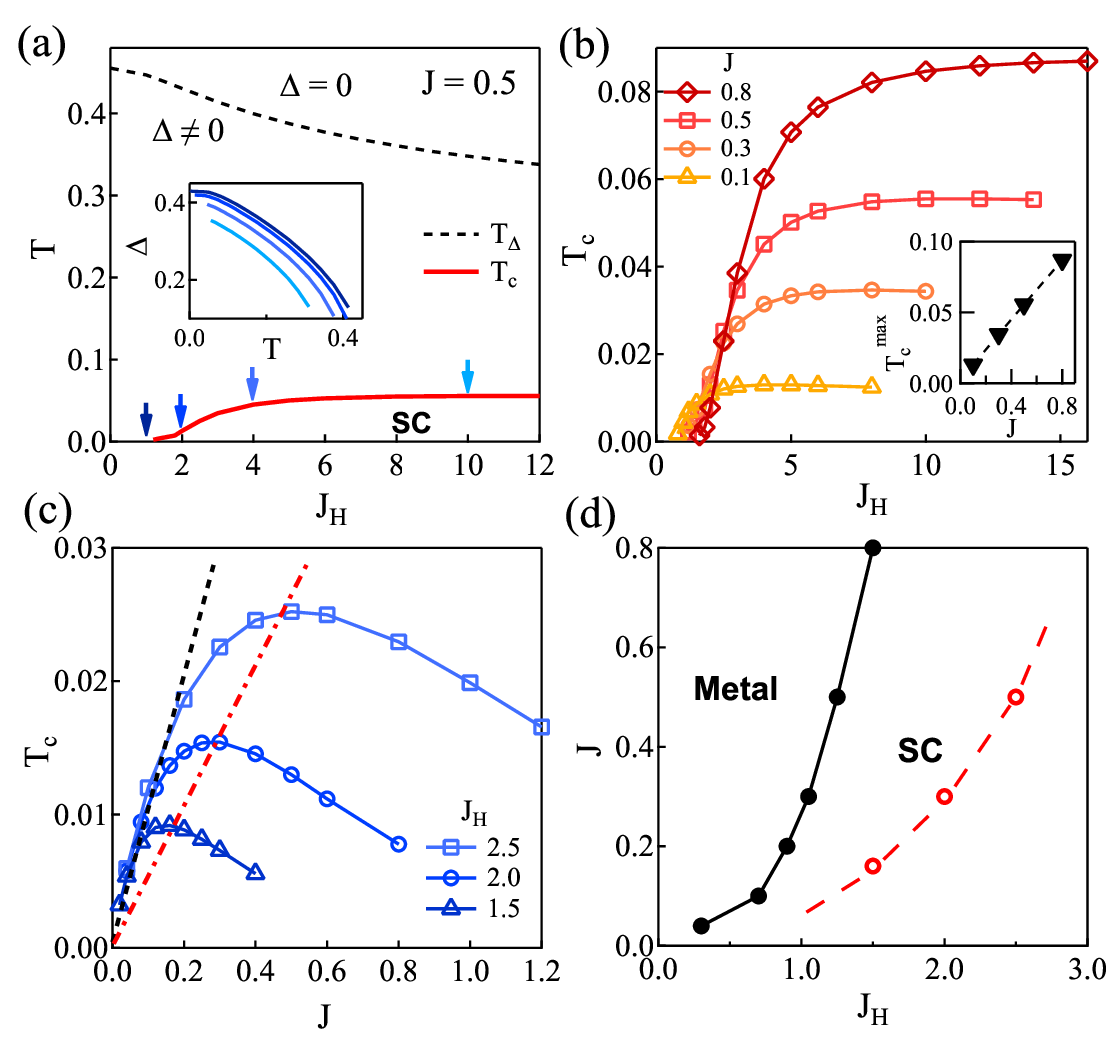}
	\end{center}
	\caption{(a) Finite temperature phase diagram of the $t$-$J_H$-$J$ model at $J=0.5$. The dashed line is  the spinon pairing temperature  $T_\Delta$ and the red line is the superconducting transition temperature $T_c$. The inset shows the temperature dependence of $\Delta$ at four different values of $J_H$ indicated by the blue arrows. (b) $T_c$ as a function of $J_H$ for different $J$. The inset shows a linear relation between maximal (saturated) transition temperature and the interlayer superexchange interaction, $T_c^\text{max}\approx 0.1 J$.  (c) $T_c$ as a function of $J$ for different $J_H$. The black dashed line shows the maximal ratio $T_c/J=0.1$, while the red dash-dotted line shows the ratio between the optimal $T_c$ and $J$, $T_c^\text{opt}/J^\text{opt}\approx 0.05$. (d) The ground state phase diagram in terms of $J$ and $J_H$. The solid line is the phase boundary between the metallic and superconducting phases and the dashed curve with red circles marks $J^\text{opt}$.}
	\label{fig:Fig2}
\end{figure}

\textit{Phase diagram---}Figure \ref{fig:Fig2}(a) shows the finite-temperature phase diagram derived for a fixed $J=0.5$. For simplicity, we set $t=1$ as the energy unit and fix $\mu=-1.44$ so that the $d_{x^2-y^2}$ electrons are quarter filled at $J_H=0$. As the temperature decreases, the $d_{z^2}$-spins first form interlayer spin singlets with a nonzero $\Delta$ below the temperature $T_\Delta$, while superconductivity emerges at a much lower temperature $T_c$. At $J_H=0$, we obtain analytically $T_\Delta=J/\ln3\approx 0.91J$. Increasing $J_H$ gradually suppresses $T_\Delta$, because the Hund's coupling term does not commute with the superexchange term, hence introduces quantum fluctuations of the spinon pairs and reduces $T_\Delta$. By contrast, $T_c$ only becomes nonzero when $J_H$  reaches some critical value $J_H^c$ ($\approx 1.2$ for $J=0.5$), then increases rapidly with $J_H$, and eventually saturates. The inset of Fig. \ref{fig:Fig2}(a) shows the temperature dependence of $\Delta$ for different $J_H$. The absence of superconductivity at small $J_H$, though with a large $d_{z^2}$ pairing amplitude, indicates that a finite Hund's coupling is necessary to transmit the interlayer pairing to $d_{x^2-y^2}$ electrons. If $J_H$ is too small, the pairing vertex is suppressed by the $|\chi_{li}|^2$ term in Eq. (\ref{eq:Hv}) and cannot support the $d_{x^2-y^2}$ pairing.

Figure \ref{fig:Fig2}(b) plots $T_c$ as a function of $J_H$ for different values of $J$. Increasing $J$ slightly increases the critical Hund's coupling, suggesting weak competition between $J$ and $J_H$. Beyond the critical $J_H$, the rapid increase of $T_c$ indicates the pairing is now effectively transmitted from $d_{z^2}$ to $d_{x^2-y^2}$ electrons. The maximal  transition temperature $T_c^\text{max}$ at large $J_H$ follows a linear relation with the interlayer superexchange, $T_c^\text{max}/J\approx 0.1$, as shown in the inset of Fig. \ref{fig:Fig2}(b). Fig. \ref{fig:Fig2}(c) plots $T_c$ as a function of $J$ for different $J_H$. $T_c$ indeed increases almost linearly at small $J$, reaching its maximum $T_c^\text{opt}$ at some optimal superexchange $J^\text{opt}$, and then starts to decrease. Interestingly, the ratio $T_c^\text{opt}/J^\text{opt}$ is around $0.05$ for all $J_H$, as shown by the red dash-dotted line. For comparison, previous Schwinger boson calculation of the $t$-$V$-$J$ model gives a much larger ratio $T_c^\text{max}/J\approx T_c^\text{opt}/J^\text{opt}\approx 0.17$ \cite{Wang2025}. The much lower  $T_c^\text{max}/J$ and $T_c^\text{opt}/J^\text{opt}$ for the $t$-$J_H$-$J$ model compared to that of the $t$-$V$-$J$ model using the same Schwinger boson approach indicates that the Hund's coupling is less effective in transmitting the pairing interaction than the hybridization, possibly because the Hund-induced holon field lacks intrinsic dynamics, which leads to a small holon quasiparticle weight and suppresses the pairing vertex. Note that all these ratios are overestimated because of the mean-field treatment of $\Delta_i$ and $\lambda_{li}$. Comparison with the auxiliary-field Monte Carlo calculations for the $t$-$V$-$J$ model \cite{QQin2023,QQin2025} suggests a reduced $T_c^\text{max}/J$ of about 30\% of these mean-field values once phase fluctuations are considered.

Figure \ref{fig:Fig2}(d) shows the phase diagram on the $J_H$-$J$ plane, where the solid line gives the phase boundary between the normal metal and superconducting states, and the dashed line marks $J^\text{opt}$ for each $J_H$. It is evident that the critical $J_H$ increases with increasing $J$ and, for a realistic $J/t\approx 0.3$, $J_H/t$ needs to surpass 1 to induce the superconductivity and 2 to reach the maximal $T_c$.

\begin{figure}[t]
	\begin{center}
		\includegraphics[width=8.5cm]{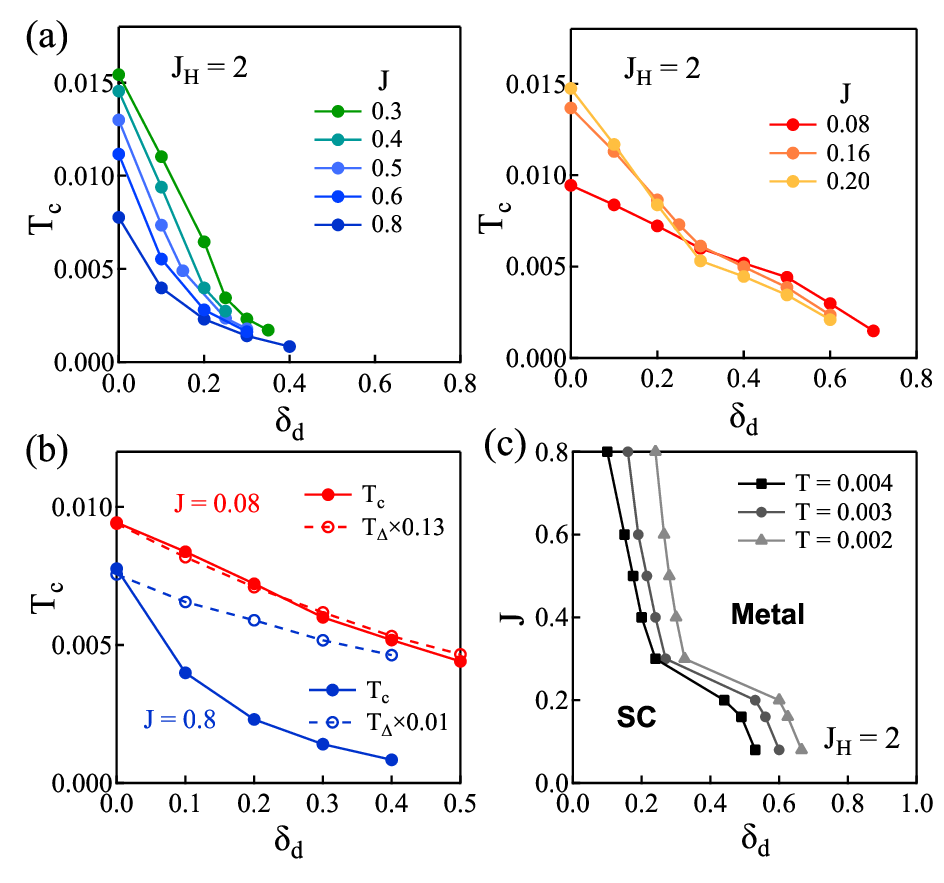}
	\end{center}
	\caption{(a) $T_c$ as a function of $d_{z^2}$ hole doping $\delta_d$ for $J_H=2$ and different values of $J$. (b) Comparison of $T_c$ and scaled $T_\Delta$ for $J=0.08$ (red) and $J=0.8$ (blue). (c) The low temperature phase diagram in terms of $J$ and $\delta_d$ at $J_H=2$. }
	\label{fig:Fig3}
\end{figure}

\textit{Effect of $d_{z^2}$ holes---}For both bulk and thin films, a key distinction is whether the $d_{z^2}$ metallization is necessary and how it may affect the superconductivity. While the Hund scenario assumes well-localized $d_{z^2}$ spins, the hybridization scenario relies on charge fluctuations (hybridization) between $d_{x^2-y^2}$ and $d_{z^2}$ orbitals and, in the large $U$ limit, requires a minimal $d_{z^2}$ hole doping to establish the superconductivity \cite{Wang2025}. Here, we study the effect of $d_{z^2}$ holes in the Hund scenario by replacing the spinon constraint $n_b=1$ with $n_b=1-\delta_d$, where  $\delta_d$ is the doped $d_{z^2}$-hole density. Figure \ref{fig:Fig3}(a) plots $T_c$ as a function of $\delta_d$ for $J_H=2$ at different $J$. We see $T_c$ decreases monotonically with increasing $d_{z^2}$ hole doping for all $J$, in stark contrast to that of the $t$-$V$-$J$ model.

Further analyses indicate that $T_c$ as a function of $\delta_d$ exhibits qualitatively different behaviors on different sides of  $J=J^\text{opt}\approx 0.3$. For $J>J^\text{opt}$, $T_c$ decreases rapidly as $\delta_d$ increases, while much slower for $J<J^\text{opt}$. To understand this, Fig. \ref{fig:Fig3}(b) compares $T_\Delta$ and $T_c$ as functions of $\delta_d$ for $J=0.08$ and $0.8$. We find $T_\Delta$ scales almost exactly with $T_c$ for $J=0.08$ but deviates strongly from $T_c$ for $J=0.8$. Thus, the suppression of $T_c$ at small $J=0.08$ is entirely due to the reduction of the spinon pairing. This is because for small $J$, the Hund's coupling is (relatively) large enough to transmit the pairing interaction to the $d_{x^2-y^2}$ electrons so that $T_c$ closely tracks the evolution of $T_\Delta$. For large $J=0.8$, when the pairing interaction cannot be effectively transmitted, $T_c$ no longer tracks $T_\Delta$ and decreases more rapidly with increasing hole doping. The resulting phase diagram is shown in Fig. \ref{fig:Fig3}(c) in terms of $J$ and $\delta_d$ at $J_H=2$. Indeed, we see an abrupt change of the phase boundary at $J=J^\text{opt}\approx 0.3$.

\begin{figure}[t]
	\begin{center}
		\includegraphics[width=8.9cm]{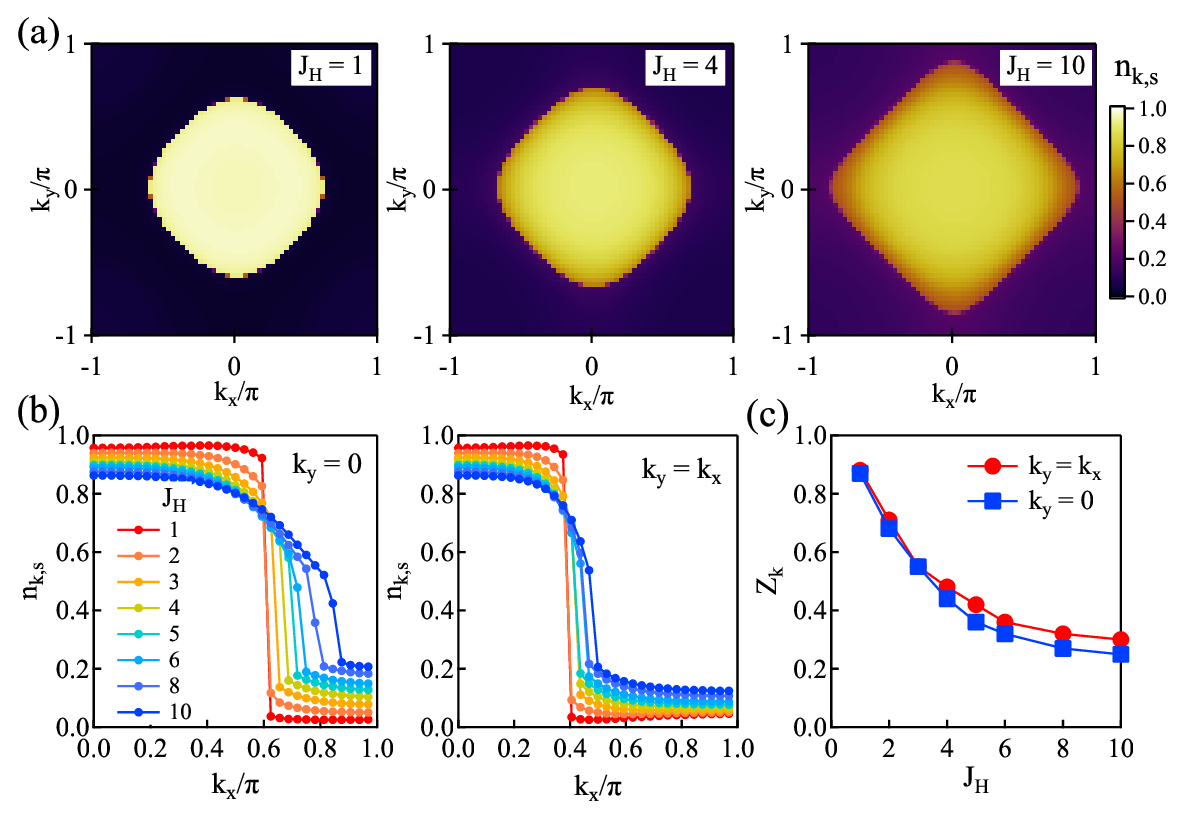}
	\end{center}
	\caption{(a) Density plot of the conduction electron occupancy per spin $n_{\mathbf{k},s}$ for $J=0.5$ and different $J_H$ obtained from the normal state solution at $T=0.0043$. (b) Variation of $n_{\mathbf{k},s}$ along $k_x$ and diagonal directions for $J=0.5$ and different $J_H$. (c) Quasiparticle weight $Z_{\bf k}$ extracted from the jump of $n_{\mathbf{k},s}$ as a function of $J_H$. Note that the $d_{x^2-y^2}$ Fermi surfaces do not split into $\alpha$ and $\beta$ sheets here because the $d_{z^2}$ electrons are assumed to be localized and the small $d_{x^2-y^2}$ interlayer hopping is neglected for simplicity.}
	\label{fig:Fig4}
\end{figure}

\textit{Normal state---}Figure \ref{fig:Fig4}(a) presents the momentum distribution of $n_{\mathbf{k},s}$, the $d_{x^2-y^2}$ electron occupancy per spin, calculated from the normal solution at low temperature for $J=0.5$ and different $J_H$. At $J_H=1$, one observes a sharp Fermi surface separating the Brillouin zone into a (nearly) doubly occupied region and an empty region, suggesting almost free $d_{x^2-y^2}$ electrons at small Hund's coupling. As $J_H$ increases, the Fermi surface is gradually enlarged (for fixed $\mu$), and $n_{\mathbf{k},s}$ on both sides of the Fermi surface deviates from its free values, $1$ and $0$, suggesting increasing electronic correlations.  However, the volume enclosed by the Fermi surface is numerically equal to the electron density $n_s=\mathcal{N}_s^{-1}\sum_\mathbf{k}n_{\mathbf{k},s}$ for all $J_H$, indicating that the Luttinger sum rule is satisfied. Fig. \ref{fig:Fig4}(b) shows the profile of $n_{\mathbf{k},s}$ along the $k_x$ axis and the zone diagonal $k_x=k_y$ for different $J_H$. There is always a sharp jump in $n_{\mathbf{k},s}$ across the Fermi  wave vector along both directions, as expected for a Fermi liquid. The size of the jump corresponds to the quasiparticle weight $Z_\mathbf{k}$. As shown in Fig. \ref{fig:Fig4}(c), $Z_{\mathbf k}$ decreases monotonically and tends to saturate as $J_H$ increases, indicating a Fermi liquid normal state for arbitrarily large Hund's coupling.

We have also calculated the density of states and the imaginary part of the self energy for $d_{x^2-y^2}$ electrons, and found neither pseudogap nor strange metal behaviors \cite{supp}. This is in stark contrast to the $t$-$V$-$J$ model, where both pseudogap feature and linear-in-$T$ scattering rate are predicted in the normal state. We attribute these distinction to the different low-energy behavior of the Hund's coupling under renormalization group. For localized  $d_{z^2}$ electrons, the inter-orbital hybridization turns into an antiferromagnetic Kondo-type scattering term, which is typically enhanced under renormalization group and causes NFL behaviors in the normal state. By contrast, the local Hund's coupling corresponds to a ferromagnetic Kondo term, which is known to flow to a weak coupling fixed point \cite{Anderson1970}, leading to a weakly correlated Fermi liquid ground state. The Hund-driven superconductivity may then be regarded as the Cooper instability of such a correlated Fermi liquid.

\textit{Discussion---}As summarized in Table \ref{tab1}, our calculations using the same approach for the bilayer two-orbital $t$-$J_H$-$J$ and $t$-$V$-$J$ models reveal several fundamental differences of the Hund-driven and hybridization-driven superconductivity in the bilayer nickelates. We find that the Hund's coupling cannot explain the NFL behavior with linear-in-$T$ resistivity observed in bulk materials under high pressure. In thin films, recent experiments suggest an anisotropic $s^{\pm}$-wave superconductivity with gap minima along the zone diagonal on $\beta$ band and a nearly constant gap on $\gamma$ band \cite{HHWen2025,ZYChen2025}. At first glance, the weak anisotropy, reduced maximum $T_c$, the absence of $\gamma$ pocket in some experiment, and the Fermi liquid normal state seem to support the Hund scenario with localized $d_{z^2}$ moments. However, the presence of $\alpha$ and $\beta$ Fermi surfaces can only arise from hybridization with $d_{z^2}$ quasiparticle bands, since the interlayer hopping of $d_{x^2-y^2}$ alone is too small to induce such a large splitting. This is the case even in the absence of the $\gamma$ pocket, which does not necessarily imply fully localized $d_{z^2}$ electrons but may indicate filled bonding bands of $d_{z^2}$ quasiparticles. Moreover, $J$ has been estimated to be reduced by about half in thin films \cite{DXYao2025film}. These together may result in a much reduced maximum $T_c$ in the Hund-driven mechanism. Thus, it seems that the Hund scenario alone is insufficient to explain either bulk or thin film nickelates, while the hybridization scenario can provide a natural and unified explanation of all experimental observations including the FL and NFL normal state, the superconducting gap anisotropy, and the maximum $T_c$. However, the two mechanisms may well cooperate to promote the interlayer pairing superconductivity. We suggest future measurements of $J$ and the true gap structures (excluding disorder effect) for more elaborate examinations.
  
\textit{Acknowledgements---}J.W. was supported by the Young Scientists Fund of the National Natural Science Foundation of China (Grants No. 12304174). Y.Y.  was supported by the National Natural Science Foundation of China (No. 12474136) and the National Key R\&D Program of China (Grant No. 2022YFA1402203). \\

\end{document}


\title{Fermi liquid and isotropic superconductivity of Hund scenario for bilayer nickelates\\
	\vspace{0.2cm}
	- Supplemental Material -}
\author{Jiangfan Wang}
\email[]{jfwang@hznu.edu.cn}
\affiliation{School of Physics, Hangzhou Normal University, Hangzhou, Zhejiang 311121, China}
\author{Yi-feng Yang}
\email[]{yifeng@iphy.ac.cn}
\affiliation{Beijing National Laboratory for Condensed Matter Physics and Institute of
Physics, Chinese Academy of Sciences, Beijing 100190, China}
\affiliation{School of Physical Sciences, University of Chinese Academy of Sciences, Beijing 100049, China}
\affiliation{Songshan Lake Materials Laboratory, Dongguan, Guangdong 523808, China}
\date{\today}

\maketitle

In this supplemental material, we provide additional derivations and figures that may help the readers.

\subsection{I. Dynamic Schwinger Boson Approach}

Using the Schwinger boson representation of the $d_{z^2}$ spins, we can derive the following action for the $t$-$J_H$-$J$ model,
\begin{eqnarray}
	S&=&\sum_{lks}\bar{c}_{lks}(-i\omega_n+\epsilon_\mathbf{k})c_{lks}+\sum_{lqs}\bar{b}_{lqs}(-i\nu_m+\lambda)b_{lqs}+\bar{\Delta}\sum_{qs}sb_{1qs}b_{2,-q,-s}+c.c.  \notag \\
	&&-\frac{2}{J_H}\sum_{lk}|\chi_{lk}|^2+\frac{1}{\sqrt{\beta\mathcal{N}_s}}\sum_{lskq}\bar{b}_{lqs}c_{lks}\chi_{l,q-k}+c.c. +2\beta\mathcal{N}_s\left(\frac{|\Delta|^2}{J}-\lambda\right), \label{eq:S}
\end{eqnarray}
where $\epsilon_{\mathbf{k}}=-2t(\cos k_x+\cos k_y)-\mu+J_H/4$, and $k\equiv(\mathbf{k},i\omega_n)$, $q\equiv(\mathbf{q}, i\nu_m)$. The constant $J_H/4$ in the electron dispersion relation comes from the decomposition of the Hund term (see Eq. (3) of the main article), which will be canceled out by the same constant in its self-energy.  Using the Dyson-Schwinger equations \cite{Wang2020} or Luttinger-Ward functional \cite{Coleman2005}, we derive the following self-energies for the electron, holon and spinon:
\begin{eqnarray}
	\Sigma_c(i\omega_n)&=&\frac{1}{\beta}\sum_{m}G_\chi(i\omega_m)G_b(i\omega_m+i\omega_n), \notag \\
	\Sigma_{\chi}(i\omega_n)&=&\frac{2}{\beta}\sum_m G_c(i\omega_m)G_b(i\omega_m+i\omega_n), \label{eq:SE} \\
	\Sigma_b(i\nu_n)&=&-\frac{1}{\beta}\sum_{m}G_c(i\omega_m)G_\chi(i\nu_n-i\omega_m). \notag
\end{eqnarray}
The Green's functions depend self-consistently on the above self-energies,
\begin{eqnarray}
	G_c(i\omega_n)&=&\frac{1}{\mathcal{N}_s}\sum_\mathbf{k}G_c(\mathbf{k},i\omega_n)=\frac{1}{\mathcal{N}_s}\sum_\mathbf{k}\frac{1}{i\omega_n-\epsilon_\mathbf{k}-\Sigma_c(i\omega_n)},\notag\\
	G_\chi(i\omega_n)&=&\frac{1}{2/J_H-\Sigma_\chi(i\omega_n)},\label{eq:G} \\
	G_b(i\nu_n)&=&\frac{\gamma_b(-i\nu_n)}{\gamma_b(i\nu_n)\gamma_b(-i\nu_n)-|\Delta|^2}, \notag
\end{eqnarray}
where $\gamma_b(i\nu_n)=i\nu_n-\lambda-\Sigma_b(i\nu_n)$. One can also derive the self-energies by first using the standard  one-loop perturbation theory, then replacing the bare Green's functions with the dressed ones, corresponding to a resummation of the ``non-crossing'' diagrams. By minimizing the free energy $\mathcal{F}=-T\ln \mathcal{Z}=-T\ln \int \mathcal{D}[c,\chi,b]\exp\{-S\}$ with respect to $\Delta$ and $\lambda$, we obtain two mean-field equations:
\begin{eqnarray}
	\frac{1}{J}&=&\frac{1}{\beta}\sum_n\frac{1}{\gamma_b(i\nu_n)\gamma_b(-i\nu_n)-|\Delta|^2},\notag\\
	1&=&-\frac{2}{\beta}\sum_n G_b(i\nu_n).
	\label{eq:MF}
\end{eqnarray}

Since the holon field does not possess intrinsic dynamics, its Green's function approaches $J_H/2$ in the limit $|\omega_n|\rightarrow \infty$. In order to perform the frequency sums, one needs to subtract this constant from $G_\chi(i\omega_n)$ by defining
\begin{eqnarray}
	g_\chi(i\omega_n)=G_\chi(i\omega_n)-\frac{J_H}{2},
\end{eqnarray}
which leads to
\begin{eqnarray}
	\Sigma_c(i\omega_n)&=&\frac{1}{\beta}\sum_{m}g_\chi(i\omega_m)G_b(i\omega_n+i\omega_m)-\frac{J_H}{4},\notag\\
	\Sigma_b(i\nu_n)&=&-\frac{1}{\beta}\sum_m G_c(i\omega_m)g_\chi(i\nu_n-i\omega_m)-\frac{J_H}{4}n_c. \label{eq:Sigma}
\end{eqnarray}
Substituting Eq. (\ref{eq:Sigma}) into $G_c(\mathbf{k},i\omega_n)$, one finds that the constant $J_H/4$ from $\epsilon_{\mathbf{k}}$ and $\Sigma_c(i\omega_n)$ cancel each other. 

\textit{$J_H=0$ limit.---}In this limit, the model can be solved analytically on the mean-field level, since all the self-energies vanish. In particular, one can obtain $T_\Delta$ and $\Delta(T=0)$, both of which are proportional to the interlayer superexchange $J$. To determine $T_\Delta$, we solve the mean-field equations (\ref{eq:MF}) with $\Delta=0$, 
\begin{eqnarray}
	\frac{1}{J}&=&\frac{1}{\beta}\sum_n\frac{1}{(i\nu_n-\lambda)(-i\nu_n-\lambda)}=\frac{1}{2\lambda}\left(\frac{2}{e^{\beta\lambda}-1}+1\right),\notag\\
	1&=&-\frac{2}{\beta}\sum_n\frac{1}{i\nu_n-\lambda}=\frac{2}{e^{\beta\lambda}-1},
\end{eqnarray}
leading to $\lambda=J$ and $1/(\beta J)=T_\Delta/J=1/\ln 3\approx 0.91 $. To determine $\Delta(T=0)$, we solve Eq. (\ref{eq:MF}) at zero temperature by transforming the frequency sums  into integrals, which gives $\lambda=J$ and $\Delta(T=0)=\frac{\sqrt{3}}{2}J\approx 0.866 J$.

\begin{figure}
	\centering\includegraphics[scale=0.6]{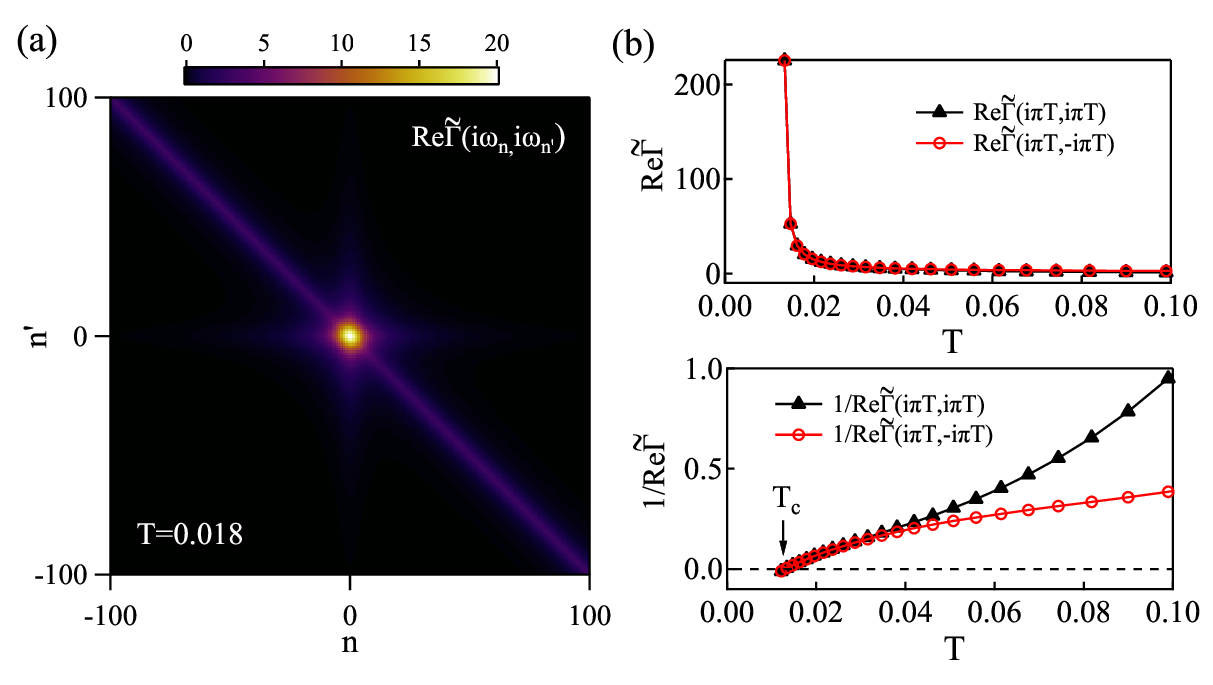}
	\caption{(a) Density plot of the real part of the pairing vertex for $d_{x^2-y^2}$ electrons, $\text{Re}\tilde{\Gamma}(i\omega_n,i\omega_{n'})$, at $J=0.5$, $J_H=2$ and $T=0.018$. (b) Temperature evolution of Re$\tilde{\Gamma}(i\pi T,\pm i\pi T)$ and its inverse at $J=0.5$ and $J_H=2$.}
	\label{figS1}
\end{figure}

\subsection{II. Cooper Instability}

The Cooper instability of $d_{x^2-y^2}$ electrons can be studied via its interlayer pairing vertex, which satisfies the following Bethe-Salpeter (BS) equation \cite{Simons2010}, 
\begin{eqnarray}
	\Gamma_{ss'}(k,k',q)=\Gamma_{ss'}^0(k,k',q)+\frac{1}{\beta\mathcal{N}_s}\sum_{k''s''}\Gamma_{ss''}^0(k,-k''-q,q)G_c(k''+q)G_c(-k'')\Gamma_{s''s'}(k'',k',q).
\end{eqnarray}
Due to the local Hund's coupling, the bare pairing vertex $\Gamma_{ss'}^0(k,k',q)=ss'\tilde{\Gamma}^0(i\omega_n,i\omega_{n'},i\nu_l)$ is momentum independent, whose expression is given by Eq. (4) of the main article. This also leads to a simplified expression for the total pairing vertex,  $\Gamma_{ss'}(k,k',q)=ss'\tilde{\Gamma}(i\omega_n,i\omega_{n'},\mathbf{q},i\nu_l)$, which is independent of the momenta $\mathbf{k}$ and $\mathbf{k}'$. Considering the static uniform solution ($q=0$), the BS equation reduces to the following linear equation,
\begin{eqnarray}
	\sum_{m}\left[\delta_{nm}-\tilde{\Gamma}^0(i\omega_n,-i\omega_{m})\frac{2}{\beta\mathcal{N}_s}\sum_{\bf k}G_c(\mathbf{k},i\omega_m)G_c(-\mathbf{k},-i\omega_m)\right]\tilde{\Gamma}(i\omega_m,i\omega_{n'})=\tilde{\Gamma}^0(i\omega_n,i\omega_{n'}),
\end{eqnarray}
which can be numerically inversed and yields $\tilde{\Gamma}(i\omega_n,i\omega_{n'})$. In our numerical calculations, the range of Matsubara frequency is chosen as $\omega_n\in\left[-(M-1)\pi T,-(M-3)\pi T,\cdots,(M-1)\pi T\right]$ with $M=2048$. Figure \ref{figS1}(a) shows the density plot of the real part of pairing vertex,  $\text{Re}\tilde{\Gamma}(i\omega_n,i\omega_{n'})$, which displays a peak at the minimal Matsubara frequency $|\omega_n|=|\omega_{n'}|=\pi T$. The imaginary part of the pairing vertex is found to be negligible compared to its real part. In addition, we also find a ridge along the $\omega_{n'}=-\omega_n$ direction, which is due to the nontrivial asymptotic behavior $G_\chi(|\omega_n|\rightarrow\infty)=J_H/2$. By replacing $G_\chi(i\omega_n)$ with its limiting value $J_H/2$, we obtain a constant pairing vertex along the $\omega_{n'}=-\omega_n$ direction:
\begin{eqnarray}
	\tilde{\Gamma}^0(i\omega_n,-i\omega_n)=\frac{J_H^2|\Delta|^2}{4\beta}\sum_m\frac{1}{\left[\gamma_b(i\nu_m)\gamma_b(-i\nu_m)-|\Delta|^2\right]^2}.
\end{eqnarray} 
However, the height of this ridge does not vary too much as temperature decreases, hence is not responsible for the superconductivity. Instead, the peak at the center (Re$\tilde{\Gamma}(i\pi T,\pm i\pi T)$) increases with decreasing temperature, and finally diverges and causes the Cooper instability, as shown in Fig. \ref{figS1}(b).

\subsection{III. Density of states above $T_c$}

\begin{figure}
	\centering\includegraphics[scale=0.65]{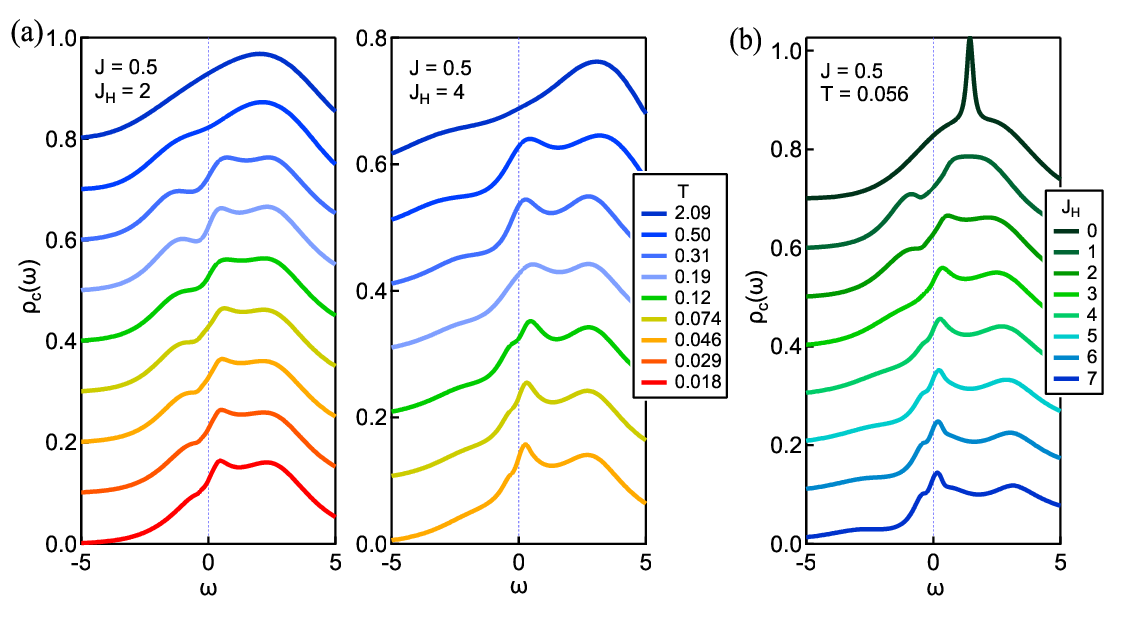}
	\caption{(a) Temperature evolution of the DOS of $d_{x^2-y^2}$ electrons above $T_c$ for $J=0.5$ and different $J_H$. (b) Evolution of the DOS with $J_H$ for fixed $J=0.5$ and temperature $T=0.056$. For clarity, the data at different temperature (or $J_H$) are shifted by a constant. }
	\label{figS2}
\end{figure}

We calculate the density of states (DOS) of $d_{x^2-y^2}$ electrons using the maximum entropy analytical continuation. Figure \ref{figS2}(a) shows the temperature evolution of the DOS above $T_c$ at $J=0.5$ for different $J_H$. One finds no pseudogap feature around the Fermi energy. Instead, we observe two peaks gradually emerging as temperature decreases, one of which locates slightly above the Fermi energy and gets more sharp at low temperatures. As shown in Fig. \ref{figS2}(b), this peak becomes more prominent and closer to the Fermi energy as $J_H$ increases, which may partly account for the increase of $T_c$ with $J_H$.     

